\documentclass[aps,preprint,preprintnumbers,amsmath,amssymb,nofootinbib]{revtex4}
\usepackage{graphicx}
\usepackage{epsfig}
\newcommand {\beg}{\begin{equation}}
\newcommand {\en}{\end{equation}}
\newcommand {\bega}{\begin{eqnarray}}
\newcommand {\ena}{\end{eqnarray}}
\begin{document}
\title{Quantized gravito-magnetic charges from WIMT: cosmological consequences}
\author{$^{1,2}$
Jes\'us Mart\'{\i}n Romero\footnote{E-mail address:
jesusromero@conicet.gov.ar}, $^{1,2}$ Mauricio Bellini
\footnote{E-mail address: mbellini@mdp.edu.ar} }
\address{$^1$ Departamento de F\'isica, Facultad de Ciencias Exactas y
Naturales, Universidad Nacional de Mar del Plata, Funes 3350, C.P.
7600, Mar del Plata, Argentina.\\
$^2$ Instituto de Investigaciones F\'{\i}sicas de Mar del Plata (IFIMAR), \\
Consejo Nacional de Investigaciones Cient\'ificas y T\'ecnicas
(CONICET), Argentina.}

\begin{abstract}
Using the formalism of Weitzenb\"ock induced matter theory (WIMT)
we calculate the gravito-magnetic charge on a topological string
which is induced through a foliation on a five-dimensional (5D)
gravito-electromagnetic vacuum defined on a 5D Ricci-flat metric,
which produces a symmetry breaking on an axis. We obtain the
resonant result that the quantized charges are induced on the
effective four-dimensional hypersurface. This quantization describes the
behavior of a test gravito-electric charge in the vicinity of a
point gravito-magnetic monopole, both geometrically induced from a
5D vacuum. We demonstrate how gravito-magnetic monopoles would
decrease exponentially during the inflationary expansion of the
universe.
\end{abstract}
\maketitle
\section{Introduction and Motivation}

For some decades, topological defects have been a very
important subject of research\cite{1}. The existence of stable
topological defect solutions was established in realistic
renormalizable theories and many developments were required in the
understanding of phase transitions. In this framework defect solutions were discovered in Higgs and Yang-Mills theories,
the Nielsen-Olesen vortex-line\cite{1'} and the t'Hooft-Polyakov
magnetic monopole\cite{2}. For instance, strings quantization has been studied
in the framework of a $AdS_5 \times S^5$\cite{4'}. The study of
the cosmological implications of topological defects has become an
area of sustained interest\cite{5}. In this context cosmic strings
may provide a viable spectrum of galaxy formation\cite{3,4}.
Another topic of great interest is the study of the evolution of magnetic monopoles in the universe.
The Big Bang cosmology predicts that a very large number of heavy, stable "magnetic monopoles" should have been produced in the early universe. However, magnetic monopoles have never been observed, so if they exist at all, they are much rarer than the Big Bang theory predicts. Inflationary cosmology resolves this problem\cite{inf}. During inflation, the density of monopoles drops exponentially, so their abundance drops to undetectable levels. Monopoles are just created before (or during) inflation, so that the rapid accelerated expansion thereafter dilutes their density to very low levels. Monopoles are created, they are created at a density of order of one per Hubble volume, which means that there is one (or fewer) in each observable Universe, so monopoles are separated by a distances of the order of the Hubble length (or larger).

However, there is a
more interesting possibility that
arises from a theory that unifies conceptually electrodynamics with a theory of gravity.
Gravito-electrodynamics was first outlined in
2006 in a cosmological context as
gravito-electromagnetic inflation\cite{mb} with the aim of describing in an unified manner
both, primordial gravitational and electromagnetic effects in the
early inflationary universe\cite{folcomp,membiela}. This is a
gravito-electrodynamic formalism constructed with a penta-vector
with components $A^b$ that can be applied to any physical system
in the framework of the induced matter theory (IMT)\cite{IMT}. This theory
is based on the assumption that ordinary matter and physical fields that we can observe in
our four-dimensional (4D) universe can be geometrically induced from a five-dimensional (5D) Ricci-flat
metric with a space-like noncompact extra dimension on which we
define a physical vacuum. Because we shall be dealing with a penta-vector
$A$ in a relativistic framework, it will be natural to consider a 5D relativistic theory.
This theory is based on the assumption
that ordinary matter and physical fields that we can observe in
our 4D universe can be geometrically induced from a 5D Ricci-flat
metric with a space-like noncompact extra dimension on which we
define a physical vacuum\cite{IMT}. The Campbell-Magaard
theorem\cite{campbell,campbellb,campbellc,campbelld} serves as a
ladder to go between manifolds whose dimensionality differs by
one. This theorem, which is valid in any number of dimensions,
implies that every solution of the 4D Einstein equations with
arbitrary energy momentum tensor can be embedded, at least
locally, in a solution of the 5D Einstein field equations in
vacuum. Because of this, the stress-energy may be a 4D
manifestation of the embedding geometry. An extension of the IMT
was realized recently using the Weitzenb\"ock induced matter
theory (WIMT)\cite{ruso}. This approach makes possible a
geometrical representation of a 5D vacuum (with a zero curvature
in the Weitzenb\"ock representation), of a nonzero curvature tensor
(in the sense of the Levi-Civita representation). The WIMT formalism was introduced
with the intent to generalize the foliation's method in the IMT of gravity, because nonstatic foliations
result in very difficult calculations. With the WIMT formalism, one can replace a dynamic foliation
over a 5D Ricci-flat space (in a Levi-Civita representation), by a static foliation from a 5D curved space (in a Levi-Civita representation), on which one defines
a 5D vacuum from the point of view of a Weitzenb\"ock
representation\cite{Wei}. Once done the foliation we can go to the
representation of Levi-Civita. This procedure provides a huge versatility to make
static foliations from 5D
curved manifolds to obtain arbitrary 4D hypersurfaces.

In this work we
shall consider a 5D space-time described by the metric
$g_{ab} = e^A_{\,\,a}e^B_{\,\,b} \eta_{AB}$\footnote{We shall
denote by $\eta_{AB}$ the tensor metric in a 5D Minskowsky
spacetime.} in a 5D Weitzenb\"ock vacuum with the aim of
quantizing the gravito-magnetic charge on a topological string
which is induced through a foliation on a 5D Ricci-flat metric.
We shall use the Weitzenb\"ock representation because it makes possible the use of
the Campbell-Magaard theorem on a Weitzenb\"ock-flat 5D space.
Such a space is Weitzenb\"ock-flat in the sense that the Riemann tensor
constructed through this kind of connections is null:
$^{(W)}R^a_{bcd}=0$. However, it cannot be Riemann-flat with
respect to the Levi-Civita connections: $^{(lc)}{R}^a_{bcd}\neq
0$.

The paper is organized as follows: in Sect. II we review the
Weitzenb\"ock representation of space and the
gravito-electromagnetic theory in the framework of WIMT. In Sect. III we
explore the evolution of gravito-magnetic charges in an expanding universe and we calculate quantizate
this charge in the Dirac's sense. Finally, in Sect. IV we develop some final comments.

\section{Weitzenb\"ock representation of space and
Gravito-electromagnetic fields from WIMT}

The Riemann tensor written with the Weitzenb\"ock
representation for the space-time characterized by the metric
$g_{ab}$, is given by
\begin{eqnarray}\label{riew}
^{(W)}R^a_{bcd}
&=&\,\overrightarrow{e}_b\left(^{(W)}\Gamma^a_{dc}\right) -
\overrightarrow{e}_c\left(^{(W)}\Gamma^a_{db}\right) +
\,^{(W)}\Gamma^n_{dc} \,^{(W)}\Gamma^a_{nb} -
^{(W)}\Gamma^n_{db}\,^{(W)}\Gamma^a_{nc}- C^n_{cb}\,
^{(W)}\Gamma^a_{dn}=0, \nonumber \label{r1}
\end{eqnarray}
where $^{(W)}\Gamma^a_{bc}$ are the Weitzenb\"ock connections and
${C}^a_{bc}$ are the coefficients of structure of the base in which we write $g_{ab}$. They
can be expressed through
$C^a_{bc}=\bar{e}^a_N\overrightarrow{e_c}(e^N_b)-\bar{e}^a_N\overrightarrow{e_b}(e^N_c)=\,^{(W)}\Gamma^a_{bc}-\,^{(W)}\Gamma^a_{cb}$.
When the absence of structure of the Minkowsky spacetime
$[\eta]_{AB}= {\rm diag}\left[  1,-1,-1,-1,-1\right]$ makes the Weitzenb\"ock torsion null, both representations (Levi-Civita and
Weitzenb\"ock), are related by the expression
\begin{equation}\label{relacionwlc}
^{(W)}\Gamma^a_{bc} = ^{(lc)}{\Gamma}^a_{bc} - \,^{(W)}K^a_{bc},
\end{equation}
where, in the absence of no-metricity $g_{ab;\,c}=0$\footnote{We can assure it because we are starting the transformation from a Minkowsky metric with coefficients $\eta_{AB}$.}, the
Weitzenb\"ock contortion $^{(W)}K^a_{bc}$ being given by the
Weitzenb\"ock torsion $^{(W)}T^a_{bc}$
\begin{equation}
^{(W)}K^a_{bc}=\frac{g^{ma}}{2}\{^{(W)}T^n_{cm}g_{bn}+\,^{(W)}T^n_{bm}g_{nc}-\,^{(W)}T^n_{cb}g_{nm}\}.
\end{equation}
Equation (\ref{relacionwlc}) shows that one can pass from a Weitzenb\"ock to a Levi-Civita
representation once the contortion $^{(W)}K^a_{bc}$ is known.

We shall consider the conditions by which we can induce curvature
and currents by means of WIMT, on a 5D space-time represented by
Cartesian coordinates. The action for the gravito-electromagnetic
fields in a 5D vacuum can be written in terms of the $F_{AB}$
tensor components or in terms of the dual tensors ${\cal F}_{ABC}$
\begin{eqnarray}
\mathcal{S}& = &\int
d^5x\sqrt{|\eta|}\left[\frac{R}{16\,\pi\,G}-\frac{1}{4} F_{AB}
F^{AB}\right]\nonumber \\
& = &\int d^5x\sqrt{|\eta|}\left[\frac{R}{16\,\pi\,G}-\frac{k}{4}
{\cal F}_{ABC} {\cal F}^{ABC} \right], \label{actionfaraday}
\end{eqnarray}
where $\frac{1}{3!}\mathcal{F}_{ABC}\mathcal{F}^{ABC}
=\frac{1}{3!\,4}\varepsilon_{ABCDE}\varepsilon^{ABCNM}F^{DE}F_{NM}=F^{NM}F_{NM}$,
and $\varepsilon_{ABCDE}
\varepsilon^{ABCNM}=3!\,2!\,(\delta^N_D\delta^M_E-\delta^N_E\delta^M_D)$,
so that when $k=\frac{1}{3!}$, we obtain that both actions
describe the same physical system\cite{ruso1}. In our case, when
we use the Lorentz gauge, we deal with a 5D vacuum, so that
$R=0$. The gravito-electromagnetic dynamics, after taking
into account the Lorentz gauge: $A^B_{;\,B}=0$ in the action
(\ref{actionfaraday}), are
\begin{eqnarray}\label{dina0}
    \Box A^K&=&\eta^{BC} A^K_{\,\,\,\,\,\,;BC}=0.
\end{eqnarray}
The gravito-magnetic currents come from the solutions for the
fields (\ref{dina0}). The last equations are compatible with a
current that has its source in
\begin{eqnarray}\label{6.00-1}F^{NB}_{\,\,\,\,\,\,\,\,\,;B}=-\eta^{AN}\left[A^M\,R^B_{\,\,\,ABM}+A^B_{\,\,\, ; M}T^M_{BA}\right].
\end{eqnarray}
Using the expression $^{(lc)}\Gamma^A_{BC}=^{(W)}\Gamma^A_{BC}
+K^A_{BC}$ it is possible to obtain the following expression
between both Faraday tensors:
\begin{equation}
^{(lc)}F^{NB}=\,^{(W)}F^{NB}+\eta^{RN}A^PK^B_{\,\,\,PR}-\eta^{RB}A^PK^N_{\,\,\,PR}.
\end{equation}
The Weitzenb\"{o}ck currents are related to the Levi-Civita ones
by the expression
\begin{equation}\label{6.03}
^{(lc)(m)}J_{AB}-\,^{(W)(m)}J_{AB}=\frac{\sqrt{|\eta|}}{2}\varepsilon_{ABCDE}\frac{1}{4}M^{[CDE]},
\end{equation}
such that the antisymmetric source
$M^{[CDE]}=\eta^{CF}\eta^{DG}\eta^{EH}M_{[FGH]}$ is given by the
expression
\begin{eqnarray}
M_{[FGH]}&=&\left(A_M\,^{(W)}T^M_{[FG}\right)\,_{;\,H]}-2\,^{(W)}T^N_{[FH|}
\,^{(W)}T^M_{N|G]}A_M-2\,^{(W)}T^N_{[GH}\,^{(W)}T^M_{F]N}A_M
\nonumber
\\ &- &\,^{(W)}T^N_{[FH|}\overrightarrow{E}_N(A_{|G]})+\,^{(W)}
T^N_{[GH|}\overrightarrow{E}_N(A_{|F]})+\,^{(W)}T^N_{[FH}\overrightarrow{E}_{G]}(A_{N})\nonumber
\\
&-&\,^{(W)}T^N_{[GH}\overrightarrow{E}_{F]}(A_{N}). \label{6.04}
\end{eqnarray}
Notice that the gauge condition in the Levi-Civita representation
$^{(lc)}A^N_{;\, N}=0$, it is preserved in the Weitzenb\"{o}ck
one: $^{(W)}A^n_{;\,n}=0$.

\section{Quantized gravito-magnetic charges in an expanding universe}

To consider the evolution of a gravito-magnetic charge in an expanding universe (for instance, during
the early inflationary universe), we shall consider the vielbein given by
$E^N_n:=diag(1,a(t),a(t),a(t),1)$, defined with respect to the 5D
Minkowsky space-time $\eta_{ab}= {\bf diag}[1, -1,-1,-1,-1]$, which
is written in cartesian coordinates. Because we are in a comoving frame, the components $U^B$, of the penta-velocity will be described by
$U\equiv (1,0,0,0,0)$. In this case the base of the
tangent space $T_p(M)$ will be given by the elements $B=\{
\frac{\overrightarrow{\partial}}{\partial
t},a(t)\frac{\overrightarrow{\partial}}{\partial
x},a(t)\frac{\overrightarrow{
\partial}}{\partial y},a(t)\frac{\overrightarrow{\partial}}{\partial z},\frac{\overrightarrow{\partial}}{\partial l}\}_p$, where the
relevant non-zero structure coefficients are
$C^i_{i0}=\dot{a}(t)$. The elements of the resulting covariant
tensor metric are given by \begin{equation} \label{6.ej1} g_{ab}=
{\bf diag}[1,-a^2(t), -a^2(t), -a^2(t), -1].
\end{equation}
In order to illustrate the formalism, now we shall consider the
case where the torsion is induced through the vielbein
\begin{eqnarray}\label{6.09}\bar{e}^{n=0}_{N=0}=1,\,
\bar{e}^{n=1}_{N=1}=1,\, \bar{e}^{n=2}_{N=1} =\varepsilon
\frac{\partial \phi(x,y)}{\partial
x},\,\bar{e}^{n=2}_{N=2}=1+\varepsilon\frac{\partial
\phi(x,y)}{\partial y},\,\bar{e}^{n=3}_{N=3}=1,\,
\bar{e}^{n=4}_{N=4}=0.\end{eqnarray} This means that the effective
4D energy momentum tensor will be
\begin{equation}\label{bajada}
^{(4D)}T^{\nu}_{\mu}=\bar{e}^{n=\nu}_{N}e^M_{n=\mu}\,^{(5D)}T^{N}_{M}|_{l=l_0}.
\end{equation}
Furthermore, the effective 4D tensor metric will be
\begin{eqnarray}\nonumber[g]_{\alpha\beta}=\left(%
\begin{array}{cccc}
  1 & 0 & 0 & 0 \\
  0 & -a^{2}(t) & 0 & 0 \\
  0 & 0 & \left(-1+\varepsilon^2\left(\left(\frac{\partial \phi(x,y)}{\partial x}\right)^2-\frac{2}{\varepsilon}\frac{\partial
  \phi(x,y)}{\partial y}-\left(\frac{\partial \phi(x,y)}{\partial y}\right)^2\right)\right)a^{2}(t) & 0 \\
  0 & 0 & 0 & -a^{2}(t) \\
\end{array}%
\right),\end{eqnarray} where $\varepsilon$ is an arbitrary small
parameter. In order to make coordinated the resulting base of the
space-time, we shall make the choice $\phi(x,y)=\arctan{(y/x)}$.
The resulting effective 4D space-time $g_{\alpha\beta}$
will be twisted
\begin{equation}\label{me}
g_{\alpha\beta}=\left. e^{A}_{\,\,\alpha} e^{B}_{\,\,\beta} \, g_{AB}\right|_{l=l_0},
\end{equation}
where the vielbein $e^{A}_{\,\,\alpha}$ are the inverse of those in (\ref{6.09}): $e^{A}_{\,\,\alpha} \bar{e}^{\beta}_A = \delta^{\,\,\beta}_{\alpha}$.
The base in terms we write the metric $g_{\alpha\beta}$ in (\ref{me}) will be free of structure, but with nonzero torsion:
$^{(W)} T^{2}_{\mu\nu}=\bar{e}^{n=2}_{N=\lambda}\, {}^{(W)} T^{\lambda}_{\mu\nu}=-\varepsilon(\partial_{\mu}\partial_{\nu}-\partial_{\nu}\partial_{\mu})\phi(x,y)$.
Using the Stokes theorem on the $xy$ plane, one can see that this
calculation is compatible with a Weitzenb\"{o}ck torsion given by
$^{(W)}T^{2}_{12}=-2\,\varepsilon\pi\,\delta^{(2)}(x,y)$. Hence,
the torsion will be on a string that is located on the $z$ axis.
In this way, although the sources are null on the 5D space-time,
the foliation drives a symmetry breaking capable of inducing an
effective torsion that generates gravito-magnetic monopoles with a
volumetric density of charge $\rho_m={}^{(4D)(lc)(m)}J_0 =  \left. {}^{(5D)}\left({}^{(lc)(m)}J_{AB} \,U^B\right) \,e^A_{\,\,0} \right|_{l=l_0}$,
such that the gravito-magnetic density of charge
$\rho_m$ can be obtained using the tensor current $^{(lc)(m)}J_{AB}$ in (\ref{6.03})
\begin{equation}
^{(lc)(m)}J_{AB}=\,^{(W)(m)}J_{AB}+\frac{\sqrt{|\eta|}}{2}\varepsilon_{ABCDE}\frac{1}{4}M^{[CDE]},
\end{equation}
$M^{[CDE]}$ being given by (\ref{6.04}). Hence, one obtains
\begin{equation}\label{cargam0}
\rho_m=-4\,
A_{2,3}\,\varepsilon\pi\,\delta^{(2)}(x,y)\,g^{22}.
\end{equation}
Notice that this result is dependent of the choice for the
vielbein (\ref{6.09}), which incorporates a non-holonomic
transformation such that $y \rightarrow
y'=y+\varepsilon\,\phi(x,y)$ represents a topological defect
similar to a dislocation\cite{defectos}.

Finally, in order to close the calculation we shall study the {\bf
quantization of gravito-magnetic and gravito-electric charges}. In this sense we follow the Vilenkin and Shellard argument\cite{vilenkin}.
This leads to a result for the amplitude of a particle to go around a closed path, $\cal{A}$. The following proportionality relation
is set as \begin{equation}\label{14115}
{\cal{A}} \sim e^{i\, Q_{gm} {\oint}_{\Gamma}\, \bf{A}\cdot \bf{dx}}= e^{i\,Q_{gm} {\int}_{\Sigma} \,\bf{B}\cdot \bf{ds}},
\end{equation}
where the surface $\Sigma$ is bounded by a closed path called $\Gamma$ and $Q_{gm}$ is the gravito-magnetic charge. In this case
we apply (\ref{14115}) to gravito-magnetic and the gravito-electric charges, $Q_{gm}$ and $Q_{ge}$, and choose $\Gamma$ as a
circumference of radius $\rho$ centered on the $\mathbf{z}$-axis. Furthermore, $\Sigma$ is considered as a cylindrical
surface bounded by $\Gamma$. All the calculation was done in order to quantize the charges on the effective 4D hypersurface,
such that the gravito-magnetic field is reduced to $\mathbf{B}= \frac{Q_{gm} }{2 \pi \rho^2} \hat{\mathbf{\rho}}$ and all the
effective tensors must take a form analogous to (\ref{bajada}). Taking into account the most basic solutions of (\ref{dina0}),
associated with the zero mode of the field and the symmetry of problem, we obtain
\begin{equation}\label{integrrrr}
\oint_{\Gamma}\mathbf{A}\cdot \mathbf{d\theta}=\left[ {\frac{2(1+\varepsilon)\varepsilon}{a(t)}e^{-K_l\frac{l-l_0}{l_0}}
\int d\phi}\right]_{l=l_0} = \frac{4\pi\varepsilon (1+\varepsilon)}{a(t)},\end{equation}
because $\int d\phi=2\pi$ over a full turn. The gravito-magnetic charge fulfills the expression
\begin{equation}\label{cargam1}
Q_{gm}= \,m \left(\frac{\varepsilon (1+\varepsilon)}{a(t)}\right),
\end{equation}
where $m$ is an integer number, and (\ref{cargam1}) is qualitatively compatible with (\ref{cargam0}). In the same way,
we can work the gravito-electric induced charge
\begin{equation}\label{cargae}
Q_{ge}= n\,\left(\frac{a(t)}{\varepsilon (1+\varepsilon)}\right),
\end{equation}
where $n$ is an integer. Hence, the product of both charges complies with Dirac's law of
quantization
\begin{equation}
Q_{ge}\,Q_{gm}= (m\,n).
\end{equation}
This result is very important and shows how the product
$Q_{ge}\,Q_{gm}$, results in an invariant on an expanding
universe. If we take in mind, for instance, an early inflationary
universe with a scale factor $a(t)=a_0\, e^{H_0 t}$, it is easy to
see that gravito-magnetic charges $Q_{gm}$ will be exponentially
decreasing during the inflationary expansion of the universe\footnote{Of course one could mean in a collapsing pre-Big Bang scenario in which
the scale factor is decreasing, so that the gravito-magnetic charges $Q_{gm}$ could be increased before the Big Bang.}, meanwhile
gravito-electric charges $Q_{ge}$ will be constant for a comoving observer.

\section{Remarks}
We have used WIMT in the particular example in which the foliation reveals a topological defect in the effective
4D arrival space-time. In the example we have obtained the presence of localized gravito-magnetic charges distributed along the $\mathbf{z}$-axis.
The gravito-magnetic charges appear to be associated with the Weitzenb\"{o}ck
torsion of space-time. This torsion is located on the topological defect and induced by a non-holonomic foliation.
The gravito-magnetic charge distribution here obtained is expressed within the Levi-Civita
derivative operator, so that it is a genuine gravito-magnetic charge distribution and was finally calculated in a Riemannian geometric
construction (although its source lies in a Weitzenb\"{o}ck torsion). The quantization of charges was carried out in the
effective 4D hypersurface. Hence, the Dirac quantization describes the behavior of a test gravito-electric charge in
the vicinity of a point gravito-magnetic monopole, both geometrically induced.

A priori, in a STM theory, we can expect that the quantization of
charges can take place in 4D space because the assumption of an
empty 5D material space. Although in this case we chose to develop
an effective quantization, a 5D quantization may be obtained at
the higher space-time when $^{(5D)}R_{AB}=\lambda ^{(5D)}g_{AB}\neq
0$ and charges must exist. Then one would apply WIMT to obtain the
effective 4D charges. The fact of using only the zero mode of
the field for the effective quantization, in addition to providing
operational simplicity, can be related to the fact that
configuration of gravito-electric and gravito-magnetic charges are
comoving/static in the example here studied. We can see how the
gravito-magnetic charge decays rapidly with time, and therefore
during the inflationary epoch they should disappear for a comoving
observer, due to the accelerated expansion of the universe. Meanwhile
gravito-electric charges remain constant for a comoving observer
who employs physical coordinates, in a more general scenario the
inner product of gravito-electric and gravito-magnetic currents can
be thought of as an invariant $I$, which in the case here studied
adopts the particularly simple form: $I=Q_{ge}\,Q_{gm}$, because
there are no gravito-magnetic currents outside the mathematical
string located on the $\mathbf{z}$-axis. Otherwise we would obtain
additional terms related to currents.

\section*{Acknowledgements}

\noindent J. M. Romero and M. Bellini acknowledge CONICET
(Argentina) and UNMdP for financial support.

\end{document}